# Pseudogap in a thin film of a conventional superconductor

Benjamin Sacépé[1,2,†], Claude Chapelier[1], Tatyana I. Baturina[3], Valerii M. Vinokur[4], Mikhail R. Baklanov[5], and Marc Sanquer[1]

[1]SPSMS, UMR-E 9001, CEA-INAC/ UJF-Grenoble 1, 17 rue des martyrs, 38054 Grenoble cedex 9, France

[2]Present address: Institut Néel, CNRS & Université Joseph Fourier, BP 166, 38042 Grenoble Cedex 9, France

[3] A. V. Rzhanov Institute of Semiconductor Physics SB RAS, 13 Lavrentjev Avenue, Novosibirsk, 630090 Russia

[4]Materials Science Division, Argonne National Laboratory, Argonne, IL 60439, USA

[5]IMEC Kapeldreef 75, B-3001 Leuven, Belgium

[†]e-mail: benjamin.sacepe@grenoble.cnrs.fr

**A superconducting state is characterized by the gap in the electronic density of states which vanishes at the superconducting transition temperature $T_c$. It was discovered that in high temperature superconductors a noticeable depression in the density of states still remains even at temperatures above $T_c$; this feature being called pseudogap. Here we show that a pseudogap exists in a conventional superconductor: ultrathin titanium nitride films over a wide range of temperatures above $T_c$. Our study reveals that this pseudogap state is induced by superconducting fluctuations and favored by two-dimensionality and by the proximity to the transition to the insulating state. A general character of the observed**

**phenomenon provides a powerful tool to discriminate between fluctuations as the origin of the pseudogap state, and other contributions in the layered high temperature superconductor compounds.**

Superconductivity, i.e. the ability of certain materials to carry loss-free current, appears due to the interaction of conducting electrons with the vibration motion of atoms[1]. The mediating action of atomic vibrations induces attraction between the electrons with the opposite spins and momenta and causes them to combine into bound pairs called Cooper pairs. Being bosons, Cooper pairs condense at the lowest possible quantum state. To break a Cooper pair and remove the two resulting unbound electrons from the condensate, one has to spend the energy $2\Delta$, the quantity $\Delta$ being called the superconducting gap. This means that the gap $2\Delta$ appears in a single electron density of states (DOS). This gap in the DOS is characteristic to a superconducting state and vanishes above the superconducting transition temperature $T_c$. However, high-temperature copper oxide-based superconductors exhibit a considerable suppression in the DOS over a wide range of temperatures above $T_c$, as evidenced by nuclear magnetic resonance, infrared conductivity, neutron scattering, transport properties, specific heat, thermoelectric power, spin susceptibility, Raman spectroscopy, tunneling conductance data, and angle-resolved photoemission[2-5]. This 'pseudogap' has become the focus of the condensed matter community as it is viewed as a key to understanding the nature of high-temperature superconductivity[6].

A possible way of understanding the pseudogap phenomenon follows from the pioneering work[7] by Anderson who proposed that high-$T_c$ superconductivity stems from the copper oxides being Mott insulators well above $T_c$ (ref. 8). The same enhanced electron-electron interaction that turns the material insulating favors electron pairing. Accordingly, from this viewpoint the pseudogap reflects the massive presence of paired electrons above $T_c$ and transforms into a superconducting gap below $T_c$.

Within the standard Bardeen-Cooper-Schrieffer theory, the stable Cooper pairs are energetically advantageous at temperatures below $T_c$ (ref. 1). Short living fluctuation-induced pairs can appear even at $T > T_c$ (ref. 9,10). However, in clean bulk conventional superconductors, the range of temperatures where fluctuations are important is extremely narrow, about $\delta T \sim 10^{-12} - 10^{-14} T_c$. Effects of fluctuations are favored by disorder and by lowering the dimensionality of the system; in the context of the superconducting films the latter condition implies that the thickness of the film is less than the superconducting coherence length $\xi$. The relevance of fluctuations is quantified by the Ginzburg-Levanyuk parameter, which in two-dimensional disordered systems is $Gi = \delta T / T_c \approx (e^2 / 23\hbar) R$, the sheet resistance $R$ characterizing the degree of disorder. Accordingly, the fluctuations gain importance in high temperature copper-oxide compounds that are made up of parallel two-dimensional conducting sheets containing copper and oxygen atoms and sandwiching in between the atoms of other kinds. Note at the same time that reduced dimensionality promotes not only superconductivity but also a competing effect, the localization. Thus increasing the sheet resistance by doping can drive bulk cuprates across the superconductor-insulator transition (SIT) into an insulating state[11-14]. At the same time, underdoped superconductors exhibit an appreciable growth of the temperature range of the existence of the pseudogap upon approaching the superconductor-insulator transition, i.e. upon increase of the sheet resistance[2]. In general, analyzing a wealth of experimental data collected during the past decade we observe a remarkable similarity between the behaviors of the underdoped copper-oxides and thin disordered films of conventional superconductors near the superconductor-to-insulator transition although important differences remain such as the anti-ferromagnetic order of the cuprates deep in the insulating side of the SIT. This heralds that an understanding of pseudogap is to be gained through identifying parallels across different classes of superconductors and makes quasi-two-dimensional films of conventional superconductors the exemplary model systems for seeking the pseudogap physics due to the presence of paired electrons above $T_c$. If the pseudogap is induced by superconducting pairing above $T_c$, one would expect that two-dimensional superconducting films should exhibit it as well. In this article we show that a pseudogap indeed exists in conventional disordered superconducting films far above $T_c$. Furthermore, the quantitative analysis of the thermal evolution of the density of states demonstrates the superconducting fluctuations origin of this pseudogap.

**Results**

**Disordered thin films as a testing ground for correlated physics.** All the above motivates our study of the ultrathin (5 nm thick) titanium nitride (TiN) films in search of the pseudogap-like behavior. The coherence length in these films, $\xi = 10$ nm (ref. 15), is much larger than the film thickness, ensuring thus that they are two-dimensional superconductors. Our recent research revealed that the 5 nm thick TiN films exhibit a superconductor-insulator transition (SIT) that can be traversed either by changing the strength of disorder[16,17], the tunable measure of which is the room temperature resistance, or by the external magnetic field[15,18]. The choice of TiN films is supported by recent findings of a mid-infrared feature in their optical conductivity[19] which is reminiscent of a pseudogap-like excitation in cuprates[4].

The observed suppression of $T_c$ with increasing $R$ (Fig. 1a) signals the enhancement of Coulomb effects[20,21]. In clean superconductors the charge carriers move freely and react instantaneously to fluctuations of the electric fields eliminating Coulomb interactions on the distances exceeding the atomic spatial scales. This ideal screening permits the phonon-mediated Cooper pairing and makes the very phenomenon of superconductivity possible. Disorder impedes the motion of electrons which then lag behind the alternations in the distribution of the electric field caused by changes in their mutual spatial positions. This breaks down dynamic screening of Coulomb interactions, and the electron-electron Coulomb repulsion start to come into play. Thus the impaired screening ignites the competition between the Coulomb forces and the superconducting pairing, effectively decreasing the strength of the Cooper electron-electron attraction.

The enhanced electron-electron interactions manifest themselves not only in the observed suppression of $T_c$ in TiN films, but, further, in the insulating trend in the resistance. This is illustrated in Fig. 1b where, in order to stress that in TiN films the distinct insulating behavior is clearly seen even at room temperature, we have plotted the inverse resistance $1/R$ as function of temperature.

**Pseudogap in the density of states.** The most direct access to the gap in the electronic excitation spectrum is offered by scanning tunneling microscope (STM) probe measuring *tunneling* conductance $G$. The density of states, $\nu(\eta)$, is found as a function of the single electron excitation energy (relative to the Fermi level), $\eta = E - E_F$, and of the temperature $T$ by recording the differential conductance $G(V) = dI/dV$ of the tunneling junction between the STM tip and the sample versus its voltage bias $V$. The tunneling conductance relates to $\nu(\eta)$ via

$$G(V,T) \propto \int d\eta\, \nu(\eta,T)[-(\partial f_T/\partial \eta)(\eta - eV)], \tag{1}$$

where $f_T$ is the Fermi distribution function at temperature $T$ (ref. 5).

We probed the tunneling conductance of superconducting films TiN1, TiN2, and TiN3, with a Scanning Tunneling Microscope (STM) cooled down in a dilution refrigerator. Importantly, the transport measurements of the resistance of films TiN1, TiN2, and TiN3, shown in Fig. 1, and scanning tunneling spectroscopy, displayed in Fig. 2, were carried out during the same run. At the lowest temperatures all three samples display a gap centered at the Fermi level, corresponding to zero bias, $V = 0$, and two peaks at the edges of the gap in accord with the Bardeen-Cooper-Schrieffer theory of conventional s-wave superconductivity[1], as observed in our previous work[22]. Upon increasing temperature, the gap gets shallower and coherence peaks are depressed. However, up to the maximal temperatures achievable in our experiments, the suppression in the density of states remains substantial. Moreover, while approaching the critical disorder strength at which the disorder-driven SIT occurs, both the depth of the pseudogap and the temperature range where it exists, show the trend to increase. Remarkably, the pronounced pseudogap persists up to 6.3 K, the highest temperature achieved in the experiment, which is about $14T_c$ for the TiN3 sample. At variance, in a 100 nm thick TiN film, which is well within the three-dimensional domain with respect to superconducting properties, a flat metallic DOS is restored exactly at $T_c$ (see ref. 23).

**Superconductivity-related origin of the pseudogap.** To uncover the nature of the observed pseudogap, we juxtapose the direct STM observations of the DOS with the results of our parallel measurements of the transport conductance. Fig. 1b shows the *T*-dependence of the dimensionless conductance $g = (h/e^2 R)$ in a semi-logarithmic scale. In all samples, it decreases logarithmically upon cooling from the room temperature down to approximately 10 K as expected for the two-dimensional disordered metal described by the standard theory of quantum corrections to conductivity. This behavior reflects that predominating contributions stem from quantum interference effects, namely, weak-localization and electron-electron interaction[24-26]. In the superconducting samples the noticeable deviation from the logarithmic dependence takes place upon approaching $T_c$ (see Methods). The main contribution to the conductance upturn comes from the superconducting fluctuations (SF) (ref. 27). Their role is two-fold: in order to organize Cooper pairs, the part of the electronic states should be borrowed from the normal metal leading to a suppression of the metallic DOS. It results in an insulating trend in the temperature dependence of the resistance, which, thus, *grows* upon decreasing temperature. At the same time, fluctuation-induced Cooper pairs short-circuit electronic conductivity, and, as the superconducting fluctuations become developed enough, this shunting effect disguises the suppressing of the DOS.

A corresponding temperature evolution of the SF-induced suppression of DOS and the resulting change in the *tunneling* conductance has been discussed by Varlamov and Dorin[28] within the framework of the perturbation theory of superconducting fluctuations[27] and is described by the "double-log" temperature dependence:

$$\frac{\delta G}{G}(V=0,\varepsilon) \cong 2Gi \ln \varepsilon, \tag{2}$$

where $\varepsilon = \ln(T/T_c)$ is the reduced temperature. To relate the observed pseudogap to superconducting fluctuations, we inspect the measured temperature evolution of the differential tunneling conductance at the zero bias. As shown on Fig. 3a, the raw data follow Eq. (2) with the high accuracy over the wide reduced temperature, $\varepsilon$, range, and the slopes of $\delta G(V=0,\varepsilon)$ vs. $\ln \varepsilon$ dependences increase with disorder, i.e. with $R$. It is noteworthy that the plots do not contain any adjusting parameter: the value of $T_c$ is independently derived from the transport measurements data (see Methods and Supplementary Fig. S1). At the same time, $\delta G(V=0,\varepsilon)$ vs. $\ln \varepsilon$ dependence offers an independent method for determining $T_c$. Indeed, the slightest change in $T_c$, eliminates the linear dependence of $\delta G(V=0,\varepsilon)$ on $\ln \varepsilon$ as seen in Fig. 3b. This dependence holds over a wide temperature range only upon the proper, within the 50 mK accuracy, choice of critical temperature $T_c$ evidencing a remarkable consistency between the *local* spectroscopic measurements and the *macroscopic* transport data, irrespectively to the spatial fluctuation of the gap amplitude observed in ref. 22 (see Fig. 3c and 3d). However, it must be noticed that the maximum energy differences in gap values observed at very low temperature are only of a few tens of μeV. At higher temperature, in the pseudogap state, these energy differences are likely to be washed out and, in any case, are beyond the energetic resolution of tunneling spectroscopy due to the large thermal smearing.

Equation (2) holds over the whole domain of Gaussian fluctuations, $Gi \ll \varepsilon \ll 1$, as long as the condition $\delta G/G \ll 1$ is satisfied. At the same time, Fig. 3 shows that $\ln \varepsilon$ law describes fairly well the experimental data even beyond the perturbative regime extending to the region where the SF corrections are large, and also persists to well above $T_c$ where $\varepsilon \geq 1$. The limit, $Gi \ll 1$, determines the temperature below which one enters the critical region where the fluctuations are correlated and cannot be any longer considered as Gaussian. In Fig. 3, this corresponds to plateaus where the DOS is nearly temperature independent. As expected from the SF theory[27], both, the temperature range of this critical regime as well as the slope of the $\ln[\ln(T/T_c)]$ dependence, increase from the less resistive sample TiN1 to the more resistive one TiN3, reflecting the increase of *Gi* with disorder.

All the above indicates that the origin of the observed pseudogap in TiN film is the suppression of DOS by two-dimensional superconducting fluctuations. To conclusively rule out other possibilities, we inspect the role of Coulomb interactions which, in principle, contribute to the suppression of the DOS (ref. 24-26), resulting in the so-called Aronov-Altshuler-Lee zero-bias anomaly. The Coulomb effects lead to the logarithmic behavior of the transport conductance at temperatures not too close to $T_c$ as shown in Fig. 1b. However, re-plotting the *tunneling* conductance near $T_c$ shown in Fig. 3a as function of $\ln T$ does not reveal a linear dependence of $G$ on $\ln T$. Thus the zero-bias anomaly does not influence DOS near $T_c$, which is not unexpected: in the close vicinity of superconducting transition the SF-induced double-logarithmic divergence of the tunneling conductance is by far stronger. Incidentally, our findings show that SF not only dominate the DOS behavior at $E_F$ but also govern the transport properties at very low voltages.

## Discussion

The question that stands out through all the pseudogap studies is whether the pseudogap phenomenon is due to an exotic electronic structure of the cuprates and the related nonconventional mechanisms of superconductivity or is a generic property of a wide class of superconducting materials. As a matter of fact, Nernst effect[29] and magnetization measurements[30] in these systems have unveiled superconducting phase fluctuations whose importance is still debated [31]. Our findings reveal that the SF-dominated temperature region in the ultrathin TiN films extends to temperatures well above $T_c$ with a striking resemblance to the behavior of the copper oxides. The SF-induced pseudogap persists up to 6.3 K, the highest temperature achieved in the experiment, which is about $14T_c$ for the sample with maximally suppressed $T_c$. This demonstrates another strong parallel between high temperature superconductors and thin films of conventional superconductors, namely, the extension of the temperature range where the pseudogap state has been observed upon approaching the superconductor-insulator transition. The conclusion to be drawn is that quantitative analysis of our data establishes unambiguously the existence and origin of the pseudogap state in ultrathin films of conventional superconductors. The pseudogap there results from the quasi-two-dimensional superconducting fluctuations suppressing the density of states and is further enhanced by the proximity to the superconductor-to-insulator transition. Given that cuprates are comprised of the conducting Cu-O planes bringing thus in the two-dimensional effects, our findings offer new insight into a fascinating pseudogap state in high temperature superconductors suggesting that quasi-two-dimensional superconducting fluctuations may play a primary role in its formation. This calls for the similar quantitative analysis of the fluctuation-related phenomena in cuprates in order to evaluate the actual importance of two dimensional fluctuation effects in the pseudogap of these materials. Such an analysis may therefore become a decisive component in disentangling superconducting fluctuations effects from other possible contributions in the long-standing mystery of the pseudogap state.

## Methods

**Sample preparation.** Our samples are ultra-thin films of titanium nitride synthesized by atomic layer chemical deposition onto a Si/SiO$_2$ substrate. TiN1 is a 3.6 nm thick film deposited at 400°C while TiN2 and TiN3 are 5.0 nm thick films deposited at 350°C. TiN3 was then slightly plasma etched in order to reduce its thickness. Electron transmission and diffraction pattern revealed the films to be made of densely-packed crystallites with a typical size of 5.0 nm. Samples were patterned into Hall bridges using conventional UV lithography and plasma etching. Well below Tc, they display a disorder-induced inhomogeneous superconducting state revealed by scanning tunneling spectroscopy performed at 50 mK[22].

**Measurements.** The STM Pt/Ir tip was aligned above one of the free $500\times500$ μm$^2$ contact pads of the Hall bridge. In order to probe the local DOS, the differential conductance of the tunnel junction, $G(V) = dI/dV$, was measured by a lock-in amplifier technique with an alternative voltage of 10 μV added to the ramped bias voltage. The tunneling current was $0.5 \div 1.0$ nA for milliVolts bias voltage yielding a tunneling resistance of about $1 \div 2$ MΩ, much higher than the sheet resistance of our samples. Hence, no voltage drop across the resistive film in series with the STM junction occurs during spectroscopy. Four probe measurements of the film resistance were carried out by acquiring both voltage and current with a low frequency lock-in amplifier technique in a four terminal configuration. Transport measurements and tunneling spectroscopy were systematically carried out during the same run in a home-built STM cooled down to 50 mK in a dilution refrigerator. Temperature of the sample holder, which was weakly coupled to the dilution refrigerator, was accurately controlled by a $RuO_2$ thermometer and a resistive heater. It is worth noticing that no measurable thermal drift of the tip position occurs in our STM between the base temperature of 50 mK and the highest measured temperature of about 6.3 K.

**Contribution of quantum corrections to the conductivity and determination of $T_c$.** The fundamentals of the theory of quantum corrections to the conductivity of disordered metals can be summarized as follows[24,27,32,33]:

1. The diffusive motion of a single electron is accompanied by the quantum interference of the electron waves [the effect is referred to as the weak localization (WL)].
2. In disordered systems the electron-electron interactions (EEI) increase.

These effects result in so-called quantum corrections to the classical Drude conductivity $G_0$ yielding the following form of the total conductivity:

$$G = G_0 + \Delta G^{\mathrm{WL}} + \Delta G^{\mathrm{EEI}}. \tag{3}$$

Quantum corrections stemming from the electron-electron interaction comprise, in their turn, two parts. A correction of the first type, known as the *interaction in a diffusion channel* (ID), involves the interaction between particles with close momenta. A second type correction is the quantum correction originating from the electron-electron interaction in the Cooper channel, where the total momentum of interacting electrons is small. This correction is essential in the systems crossing over to the superconducting state and is referred to as the *fluctuational superconductivity*. The corrections to the conductivity caused by the fluctuation-induced formation of Cooper pairs or, which is the same, by the interaction in the Cooper channel branch into three distinct types. The first one is the famous Aslamazov-Larkin term (AL) describing the direct contribution of superconducting fluctuations into conductivity, i.e. the flickering short-circuiting of conductivity by the fluctuating Cooper pairs. The second type of conductivity corrections appear since in order to form the fluctuation Cooper pairs above $T_c$, some of the normal electron states have to be borrowed. This results in some depression of the normal excitations density of states (DOS), or equivalently in an effective decrease in the number of normal conducting electrons. According to the Drude formula this leads to the decrease in the normal electron conductivity. This contribution is known as the density of states suppression term. The third type correction is the Maki-Thompson correction (MT) coming from the coherent scattering on impurities of the electrons forming a Cooper pair. In terms of these corrections the total conductivity acquires the form:

$$G = G_0 + \Delta G^{\mathrm{WL}} + \Delta G^{\mathrm{ID}} + \Delta G^{\mathrm{AL}} + \Delta G^{\mathrm{DOS}} + \Delta G^{\mathrm{MT}}. \tag{4}$$

In quasi-two-dimensional disordered superconducting systems, where the thickness $d$ of a superconducting film is larger than both, the Fermi wave length, $\lambda_F$ and the mean free path $\ell$, but is less than the lengths responsible for the electron-electron interaction in both, the Cooper channel ($\xi$ - the superconducting coherence length) and the diffusion channel ($L_T = \sqrt{2\pi\hbar D/(k_B T)}$) is the thermal coherence length), i.e., $\lambda_F, \ell < d < \xi, L_T$, these corrections cast into the following:

$$\frac{\Delta G^{\mathrm{WL}}(T) + \Delta G^{\mathrm{ID}}(T)}{G_{00}} = A \cdot \ln\left[\frac{k_B T \tau}{\hbar}\right], \tag{5}$$

where $G_{00} = e^2/(2\pi^2\hbar)$.

$$\frac{\Delta G^{AL}(T)}{G_{00}} = \frac{\pi^2}{8} \cdot \left[\ln\left(\frac{T}{T_c}\right)\right]^{-1}, \tag{6}$$

$$\frac{\Delta G^{DOS}(T)}{G_{00}} = \ln\left[\frac{\ln(T_c/T)}{\ln(k_B T_c \tau/\hbar)}\right], \tag{7}$$

$$\frac{\Delta G^{MT}(T)}{G_{00}} = B \cdot \beta(T/T_c), \tag{8}$$

where $B = \ln\{2\pi\hbar/[e^2 R \ln(\pi\hbar/(e^2 R))]\}$ ($R$ is the resistance per square) and $\beta(T/T_c)$ is the electron-electron interaction strength function introduced and tabulated by Larkin[32]:

$$\beta(T/T_c) = \frac{\pi^2}{4}\sum_m (-1)^m \Gamma(|m|) - \sum_{n\geq 0}\Gamma''(2n+1),$$

$$\Gamma(m) = \left[\ln\frac{T}{T_c} + \psi\left(\frac{m}{2}+\frac{1}{2}\right) - \psi\left(\frac{1}{2}\right)\right]^{-1}. \tag{9}$$

The temperature dependencies corresponding to the corrections calculated from Eqs. (5) - (8) together with the curve describing the sum of all the corrections are shown in Supplementary Fig. S1 by solid lines. The experimental data are shown as yellow-filled circles. The alignment of the theoretical curves and the experimental data was done at the reference point 10 K according to formula:

$$R^{(i)}(T) = \frac{1}{\Delta G^{(i)}(T) + 1/R(T=10\text{ K})}. \tag{10}$$

The fitting parameters here are the coefficient $A$ from Eq. (5) and $T_c$. One immediately notes that down to the temperature $T = T_{SF}$, where $T_{SF} \approx 7.2$ K, $T_{SF} \approx 5.7$ K, and $T_{SF} \approx 2.9$ K for samples TiN1, TiN2, and TiN3, respectively, the contributions to the resistance from superconducting fluctuating corrections, $\Delta G^{AL}$, $\Delta G^{DOS}$, and $\Delta G^{MT}$, *all compensate each other*, (in the determination of these temperatures the criterion that the divergence of the curve corresponding to the sum of *all* corrections and the curve corresponding to the WL+ID contributions only was 1 %). Therefore, at temperatures above $T_{SF}$, the resistance curves follow the temperature dependencies defined by $\Delta G^{WL}(T) + \Delta G^{ID}(T)$. This is the reason why superconducting fluctuations at high temperatures are hardly detectable by conductance measurements, but can be observed by tunneling spectroscopy which probes specifically the DOS correction. At lower temperatures, $\Delta G^{MT}(T)$ dominates the temperature behavior of the resistance and the corresponding curve bends down. When fitting the data, for the *R*, entering the parameter *B*, its value at 10 K was taken. Importantly, the fitting parameters *A* and $T_c$ are determined *independently*: *A* is found from the fit in the "high"-temperature interval from predominantly WL+ID correction curve, whereas $T_c$ is determined from the fit at lower temperatures in the Maki-Thompson dominated domain. This ensures the *high precision* determination of $T_c$ presented in Supplementary Table S1.

**Acknowledgements** We thank H. Aubin, K. Behnia, L. Benfatto, H. Bouchiat, M. Feigel'man, A. Finkel'stein, Ø. Fischer, M. Houzet, V. Mineev, D. Roditchev, D. Shahar, and A. Varlamov for valuable discussions. T.B. acknowledges the support from Russian Foundation for Basic Research (Grant No. 09-02-01205), V.V. was supported by the U.S. Department of Energy Office of Science under the Contract No. DE-AC02-06CH11357.

**Author Contributions** B.S. and C.C. carried out the experiments. B.S., C.C., T.B., and V.V. analyzed the data. M.B. and T.B. prepared the TiN thin films samples. B.S, C.C., T.B., and V.V. wrote the paper. M.S. and T.B. initiated this work. All the authors discussed the results and commented on the manuscript.

**Competing Interests** The authors declare that they have no competing financial interests.

**Figure 1 Transition to a superconducting state of the TiN films.** (**a**) Sheet resistance (R) versus temperature plots of TiN films at low temperatures. As the resistance at room temperature grows, the critical superconducting temperature $T_c$ decreases (the plots for samples TiN1 to TiN3 were published in ref. 22 in a different form) and the sample TiN4 becomes insulating at low temperatures. (**b**) The same data as in **a** but extended to room temperatures and re-plotted as the dimensionless conductance $g = (h/e^2 R)$. The black straight lines highlight this behaviour. The semi-logarithmic representation reveals logarithmic decrease of the conductance with temperature due to weak localization and Coulomb interaction effects. Transition temperatures equal to 1.3 K, 1.0 K and 0.45 K for TiN1, TiN2 and TiN3, respectively, are obtained by fitting the conductance curves with all the quantum corrections to the conductivity. The fitting procedure is described in the Methods. The fits are shown as solid curves.

**Figure 2 Pseudogap in the density of states.** Three dimensional plots of the tunneling conductance $G(V, T/T_c)$ normalized by the conductance measured at high voltage and low temperature as a function of bias voltage and normalized temperature $T/T_c$ for superconducting films TiN1 (**a**), TiN2 (**b**), and TiN3 (**c**). Black lines mark the spectra measured at $T/T_c = 1, 1.5, 2, 3$ illustrating that the pseudogap state grows more pronounced and extends over the wider temperature range as disorder increases. The suppression of the density of states of the TiN3 sample remains visible up to $T = 14 T_c$.

**Figure 3 Temperature dependence of the zero bias tunneling conductance.** The thermal evolution of the normalized differential conductance

$\delta G/G = (G(V=0,\varepsilon) - G_0)/G_0$, where $G_0$ is the differential conductance at high voltage and high temperature, is represented as a function of the reduced temperature $\varepsilon = \ln(T/T_c)$ in semi-logarithmic plots. (**a**) The dashed straight lines accentuate the double-logarithmic temperature dependence (i.e. as function of $\ln[\ln(T/T_c)]$) of the pseudogap at the Fermi level for each of the three samples. (**b**) Illustration of the no-fitting-parameter character of our plots. Even the slight modification of $T_c$ (for about 50mK) would have eliminated the linear dependence of $\delta G/G$ upon $\ln[\ln(T/T_c)]$. (**c,d**) Independence of the double-logarithmic temperature behaviour upon the position at which the local superconducting gap $\Delta$ is measured at $T \ll T_c$. The spatial fluctuations of $\Delta$ observed in ref. 22 do not affect the double-logarithmic temperature behaviour as shown for TiN1 (**c**) and TiN3 (**d**). The curves are shifted for clarity.

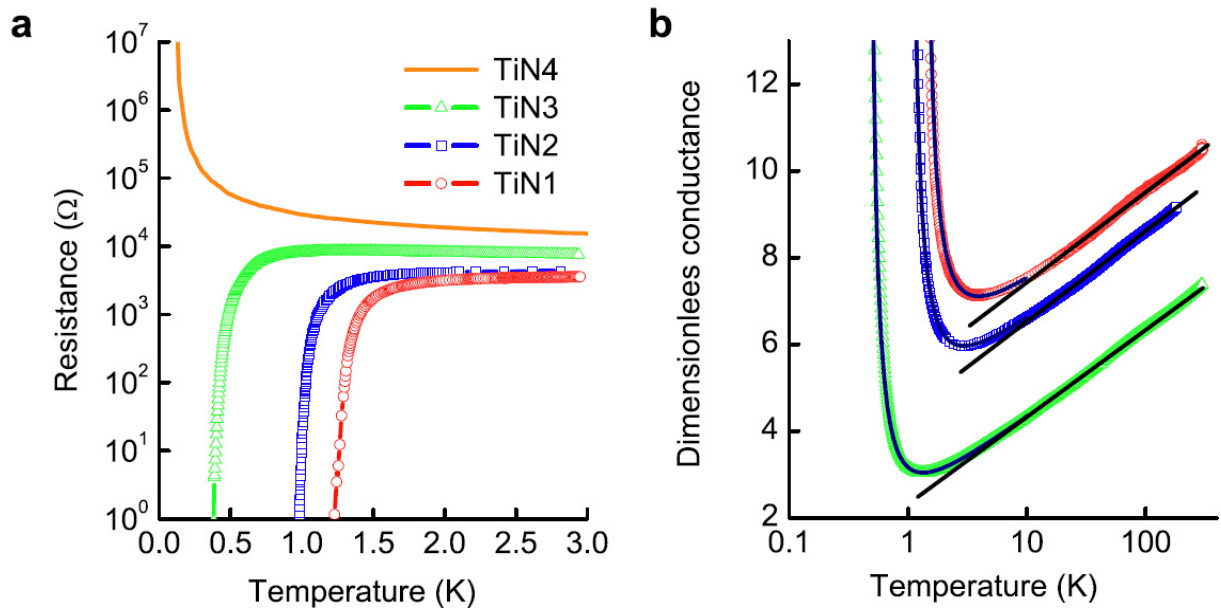

**Figure 1**

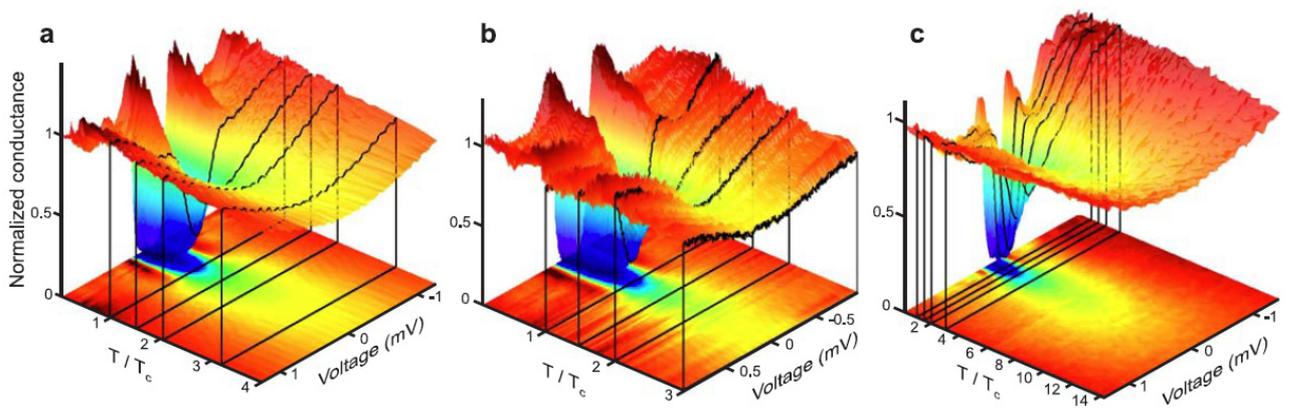

**Figure 2**

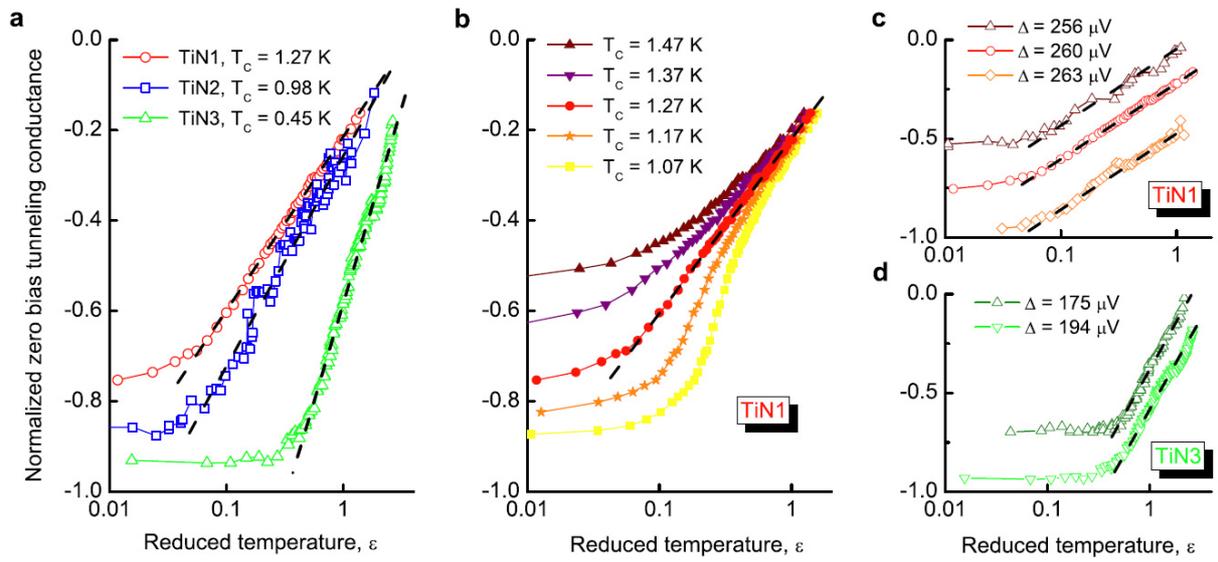

**Figure 3**

# Supplementary information

# Pseudogap in a thin film of a conventional superconductor

Benjamin Sacépé[1,2,†], Claude Chapelier[1], Tatyana I. Baturina[3], Valerii M. Vinokur[4], Mikhail R. Baklanov[5], and Marc Sanquer[1]

[1]SPSMS, UMR-E 9001, CEA-INAC/ UJF-Grenoble 1, 17 rue des martyrs, 38054 Grenoble Cedex 9, France

[2]Present address: Institut Néel, CNRS & Université Joseph Fourier, BP 166, 38042 Grenoble Cedex 9, France

[3]A. V. Rzhanov Institute of Semiconductor Physics SB RAS, 13 Lavrentjev Avenue, Novosibirsk, 630090 Russia

[4]Materials Science Division, Argonne National Laboratory, Argonne, IL 60439, USA

[5]IMEC Kapeldreef 75, B-3001 Leuven, Belgium

[†]e-mail: benjamin.sacepe@grenoble.cnrs.fr

The supplementary information includes Figure S1 and Table S1.

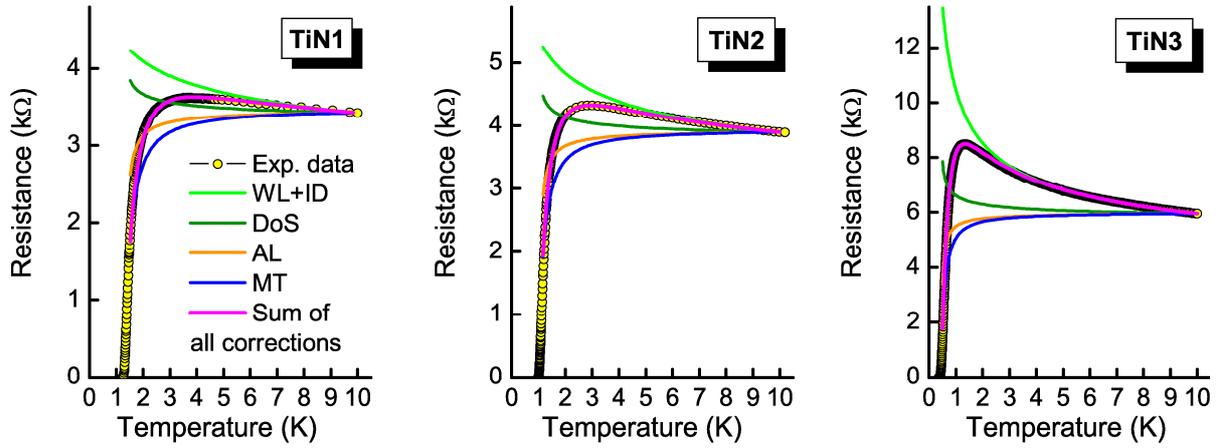

**Supplementary Figure S1 Quantum corrections to the conductivity.** Theoretical curves describing temperature behavior of quantum corrections to sheet resistance $R(T)$ and their fit to experimental data. Theoretical curves are shown by solid lines and yellow circles represent the experimental data.

**Supplementary Table S1 Fitting parameters.**

|      | $T_c$ (K) | $A$  | $B$  |
|------|-----------|------|------|
| TiN1 | 1.3       | 2.40 | 1.74 |
| TiN2 | 1.0       | 2.50 | 1.71 |
| TiN3 | 0.45      | 2.55 | 1.78 |